%% file: sample-acmtog-SIGGRAPH-submission.tex
\documentclass[acmtog]{acmart}

\acmSubmissionID{1034}
\usepackage{booktabs} 
\usepackage{bm}
\usepackage{graphicx} 
\usepackage{subcaption} 
\usepackage{multirow}
\usepackage{enumitem}

\setcopyright{rightsretained}
\acmConference[SA Conference Papers '24]{SIGGRAPH Asia Conference Papers 2024}{December 03--06, 2024}{Tokyo, Japan}
\acmYear{2024}
\copyrightyear{2024}
\acmMonth{12}
\acmISBN{979-8-4007-1131-2s/24/12}
\acmDOI{10.1145/3680528.3687683}

\acmArticle{86}

\usepackage[ruled]{algorithm2e} 

\SetAlFnt{\small}
\SetAlCapFnt{\small}
\SetAlCapNameFnt{\small}
\SetAlCapHSkip{0pt}
\acmJournal{TOG}

\begin{document}
\title{PDP: Physics-Based Character Animation via Diffusion Policy}

\author{Takara Truong}
\orcid{0000-0001-9252-2428}
\affiliation{%
 \institution{Stanford University}
 \city{Stanford}
 \state{Ca}
 \postcode{94305}
 \country{USA}}
\email{takaraet@stanford.edu}

\author{Michael Piseno}
\affiliation{%
 \institution{Stanford University}
 \city{Stanford}
 \state{CA}
 \country{USA}
}
\email{mpiseno@stanford.edu}

\author{Zhaoming Xie}
\affiliation{%
\institution{Stanford University}
\city{Stanford}
\state{CA}
\country{USA}}
\email{zxieaa@gmail.com}

\author{C. Karen Liu}
\affiliation{%
    \institution{Stanford University}
    \city{Stanford}
    \state{CA}
    \country{USA}
}
\email{karenliu@cs.stanford.edu}

\newcommand{\karen}[1]{\textcolor{red}{Karen: #1}}
\newcommand{\zhaoming}[1]{\textcolor{pink}{Zhaoming: #1}}
\newcommand{\michael}[1]{\textcolor{blue}{Michael: #1}}
\newcommand{\takara}[1]{\textcolor{blue}{Takara: #1}}


\begin{abstract}
Generating diverse and realistic human motion that can physically interact with an environment remains a challenging research area in character animation. Meanwhile, diffusion-based methods, as proposed by the robotics community, have demonstrated the ability to capture highly diverse and multi-modal skills. However, naively training a diffusion policy often results in unstable motions for high-frequency, under-actuated control tasks like bipedal locomotion due to rapidly accumulating compounding errors, pushing the agent away from optimal training trajectories.  The key idea lies in using RL policies not just for providing optimal trajectories but for providing corrective actions in sub-optimal states which gives the policy a chance to correct for errors caused by environmental stimulus, model errors, or numerical errors in simulation. Our method, Physics-Based Character Animation via Diffusion Policy (PDP), combines reinforcement learning (RL) and behavior cloning (BC) to create a robust diffusion policy for physics-based character animation. We demonstrate PDP on perturbation recovery, universal motion tracking, and physics-based text-to-motion synthesis.  

\end{abstract}


\begin{teaserfigure}
  \includegraphics[width=1.0\textwidth]{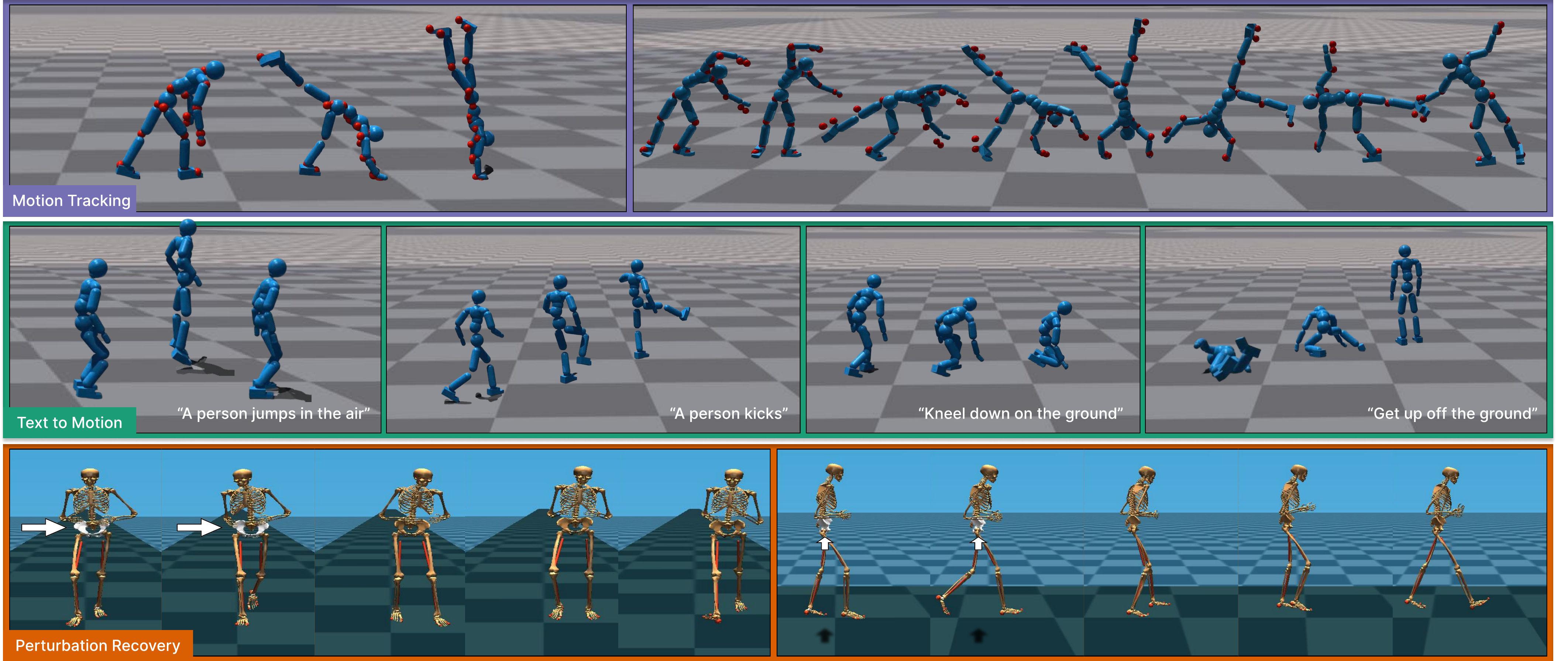}
  \centering
  \caption{PDP performs well across a diverse range of physics-based application domains. Top: Motion tracking. PDP is capable of tracking difficult and highly dynamic motions such as handstands and cartwheels. Middle: Text-to-motion. PDP is also capable of following user-provided text instructions. Bottom: Robustness to perturbations. PDP learns robust recovery strategies from random perturbations.}
  \label{fig:teaser}
\end{teaserfigure}

\maketitle

\input{introduction}

\input{related}

\input{methods}

\input{experiments}

\input{discussion}

\input{conclusion}

\bibliographystyle{ACM-Reference-Format}
\bibliography{sample-bibliography}

\input{figure_only}

\end{document}

%% file: introduction.tex

\section{Introduction}

Developing a framework capable of generating diverse human movements that can  traverse and interact with the environment is a crucial objective in character animation with broad applications in robotics, exoskeletons, virtual/augmented reality, and video games. Many of these applications demand not only a diverse range of human kinematic poses, but also the physical actions needed to achieve them. Previous works have demonstrated that physics-based control tasks can be formulated as a Markov Decision Process and solved through a Reinforcement Learning (RL) algorithm, or as a regression problem and solved using supervised learning techniques such as Behavior Cloning (BC). Despite the achievements observed in dynamic motor skills learning through both RL and BC methodologies, they encounter difficulties in effectively capturing the diversity and multi-model characteristics inherent in human motions.

To address this issue, recent works have explored various generative models. While Conditional Variational Autoencoders (C-VAEs) and Generative Adversarial Networks (GANs) have been used to capture humanoid skills, VAE-based models suffer from sensitive trade-off between diversity and robustness, while GAN-based methods often suffer from mode collapse without additional objectives \cite{dou2022case}. Although diffusion models have been used to generate varied kinematic human motion \cite{tevet2023human, tseng2022edge}, their application in high-frequency control domains is relatively unexplored. Recent work in robotics shows that BC combined with diffusion models can effectively learn diverse and multi-modal actions for real-world execution \cite{chi2023diffusionpolicy, huang2024diffuseloco}. However, naively training a diffusion-based BC policy is ineffective for physics-based character animation due to compounding errors in high-frequency, under-actuated control tasks, exacerbating the domain shift problem. This issue is especially prominent in bipedal locomotion, where accumulated errors can quickly lead to falling. Can we combine the strengths of RL and diffusion-based BC policy, such that a physically simulated character can perform a diverse set of tasks robustly against distribution shifts due to disturbances in the environment, error in model prediction, or numerical errors in simulation?

We introduce PDP, a novel method that learns a robust diffusion policy for physics-based character animation, addressing the noted challenges. PDP leverages diffusion policies \cite{song2020diffusion} and large-scale motion datasets to learn diverse and multimodal motor skills through supervised learning and diffusion models. To overcome sensitivity to domain shifts, PDP uses expert RL policies to gather physically valid sequences of observations and actions. However, using RL for data collection alone does not resolve domain shift sensitivity.

Our key insight is that RL policies provide not only optimal trajectories but more importantly corrective actions from sub-optimal states. We employ a sampling strategy from robotics literature \cite{xie2020learning}, collecting noisy-state clean-action paired trajectories to train the diffusion policy. We find that the choice of pairing noisy state with clean actions is a critical detail that contributes to producing a robust policy, outperforming the standard clean-state-clean-action trajectory collection and noisy-state-noisy-action sampling strategies for domain randomization. Additionally, we can now pool together data collected by small-task RL policies which can be efficiently trained, and leave the learning of diverse tasks on large-scale datasets to supervised learning.


PDP is a versatile method applicable to various motion synthesis tasks and agnostic to training datasets. We evaluate PDP on locomotion control under large physical perturbations, universal motion tracking, and physics-based text-to-motion synthesis, using different motion capture datasets for each application.
Our model captures the multi-modality of human push recovery behavior, outperforming VAE-based methods and deterministic multi-layer perceptron networks on the Bump'em dataset \cite{addbio}. It can also track $98.9\%$ of all AMASS motions and \cite{amass2019} generate motion from textual descriptions \cite{guo2022Humanml3d}. Our contributions are as follows:
\begin{itemize}
    \item We present a method of robust BC that scales to large motion datasets without the need for complex training architectures and can easily adapt to new skills. 
    \item We analyze the effect of different sampling strategies for data augmentation on model performance. 
    \item We introduce physics-based models that support locomotion control, motion tracking, and text-to-motion tasks.
\end{itemize}



%% file: related.tex
\section{Related Work}

{\subsection{Physics-Based Character Animation}}

In physics-based character animation, the central challenge is developing systems that learn a diverse range of realistic motions. Such motions manifest in motion tracking, motion generation, and task-oriented applications that require consideration of physical interactions.

Methods for learning individual or relatively similar motions, such as walking, running, or jumping are well-established \cite{2018-TOG-deepMimic, peng2021amp}. However, these methods often do not suffice to complete tasks that involve multiple skills. Other methods have been successful in learning motion tracking policies capable of multiple skills using model-free RL \cite{bergamin2019drecon, park2019learning} and model-based RL \cite{supertrack2021}. However, scaling to large diverse datasets is challenging. UniCon \cite{wang2020unicon} introduced a motion tracking controller that scales to large and diverse motion datasets by using a novel constrained multi-objective reward function. Other works propose a mixture of experts \cite{won2020scalable}, where different experts specialize in different skills. PHC \cite{luo2023_PHC} proposes an iterative approach to learning a large number of skills sequentially.

Another challenge in physics-based character animation is capturing diversity in motion data for use in downstream tasks. Human behaviors are multimodal, meaning a range of plausible behaviors can be employed in the same situation. A common method of capturing diversity in motions is to employ a Variational Autoencoder (VAE) to learn a latent space of skills, then sampling from the VAE prior to produce a wide range of motions \cite{Yao2022ControlVAE, yao2023moconvq, zhu2023neural, characterControlVAE, merel2018neural}. These latent motion representations can then be used for downstream tasks such as motion generation \cite{luo2024pulse} or object iterations \cite{merel2020catchcarryreusable}. Adversarial methods have also been proposed for capturing motion diversity \cite{peng2022ase, dou2022case}, which combine a diversity reward and adversarial reward that encourage the policy to mimic the motion distribution.

Robustness issues also arise in behavior cloning methods where error accumulation can easily push the policy out of distribution. One method for improving policy robustness is to continually roll out the current policy, collecting on-policy data to train a student in the next learning iteration, as in DAgger \cite{ross2011DAgger}. Alternatively, robustness can be achieved by injecting perturbations into the state-action pairs in the training dataset, effectively expanding the distribution of states seen during training, similar to DASS \cite{xie2020learning}.



\subsection{Diffusion Models for Motion Synthesis and Robotics}
Similar to VAEs, Diffusion models represent another category of generative AI and have exhibited success in the domain of kinematic motion synthesis, showcasing the capability of generating diverse and intricate human motion patterns \cite{tevet2023human, tseng2023edge}. Recently, Diffusion Policy \cite{chi2023diffusionpolicy} has effectively applied diffusion models to robotic manipulation tasks, human-robot collaborative endeavors \cite{10310116}, and tasks involving following language instructions \cite{zhang2022lad}. These models have primarily concentrated on high-level motion planning with a limited action space, such as forecasting the end-effector trajectory. While effective in low-frequency environments, the application of Diffusion Policy to high-frequency scenarios where minor inaccuracies in model predictions could result in failure, such as in physics-based character animation, remains relatively unexplored. Concurrent work, DiffuseLoco \cite{huang2024diffuseloco}, is similar to PDP, employing a diffusion model to distill an offline dataset of multimodal skills, however they focus on simple locomotion gaits due to their policy being deployed on a real robot.


%% file: methods.tex
\section{Methods}


Our method consists of three stages. First, we train a set of expert policies, each specialized in a small task but together completing a wide variety of motion tracking tasks in a physics simulator. Second, we generate state-action trajectories from the trained policies stochastically to build a dataset with noisy-state and clean-action trajectories. Lastly, we train a diffusion model via Behavior Cloning (BC) to obtain a single policy that can perform all tasks. Fig.~\ref{fig:arch} gives an overview of our system. 

\subsection{Expert Policy Training}
\label{sec:DASS}

We aim to obtain a control policy $\pi_{\text{PDP}}: \mathcal O \times \mathcal T \to \mathcal A$ to control a humanoid character, where $\mathcal O$ is the set of observations that describes the state of the character, $\mathcal T$ is the set of tasks, and $\mathcal A$ is the set of actions used to control the humanoid character. Such control policies can be trained via reinforcement learning. However, when the set $\mathcal T$ is large, it may be challenging to train a single policy to master all tasks, while it is relatively easy to train policies that specialize in a subset of tasks. We can divide the task set $\mathcal T$ into subset $\{ \mathcal T_1, \mathcal T_2, \dots, \mathcal T_k\}$, where $\bigcup_i \mathcal T_i = \mathcal T$, and train a expert policy $\pi_{\mathcal T_i}$ for each $\mathcal T_i$. The strategy for dividing the task is not critical, as long as it results in a set of policies that can generate desired state-action trajectories.

\subsection{Stochastic Data Collection}
In the second stage, we utilize the expert policies to generate a dataset for BC. For each task $\mathcal T_i$, we create a dataset $\mathcal D_{\mathcal T_i}$ by rolling out policy $\pi_{\mathcal T_i}$ and collecting trajectories. Specifically, we sample a motion task $\tau \in \mathcal T_i$, and run the policy to generate a sequence $\{\bm{o}_0, \bm{a}_0, \bm{o}_1, \bm{a}_1, \dots, \bm{o}_N, \bm{a}_N \}$, where $\bm{a}_t = \pi_{\mathcal T_i}(\bm{o}_t, \tau) + \epsilon$ is a noisy version of optimal action proposed by the expert policy. The tuples $(\bm{o}_t, \tau, \pi_{\mathcal T_i}(\bm{o}_t, \tau))$  which correspond to the observation, task/goal information, and action are added to the dataset $\mathcal D_{\mathcal T_i}$. We repeat the data collection process until a maximum number of data points are collected, and use $\mathcal D = \bigcup \mathcal D_{\mathcal T_i}$ as the dataset for BC. Note that the optimal action, not the noisy action $a_t$, is stored in $\mathcal D_{\mathcal T_i}$ and can be thought of as a corrective action from a noisy observation. This important detail, inspired by the DASS strategy proposed by \cite{xie2020learning}, results in a training set that consists of sequences of noisy-state and clean-action pairs.  This allows the collected data to cover a wider range of observation space compared to naively collecting clean optimal state-action trajectories, effectively creating a "noise band" around the clean trajectories. Our method extends DASS by further widening the noise band. Specifically, we generate short recovery episodes by initializing the character with a random root position and orientation offset from its original motion and allow it to recover to the original motion over several timesteps. This approach applies the noise band not only to the joints but also to the character's overall pose, helping to mitigate drift over time.

 


Another potential option for sampling is to collect noisy-state and noisy-action pairs, a common domain randomization practice in robotics to battle the sim-to-real gap. We found this randomization strategy produces less robust policies for character animation. 




\begin{figure}[t]
\includegraphics[width=0.5\textwidth]{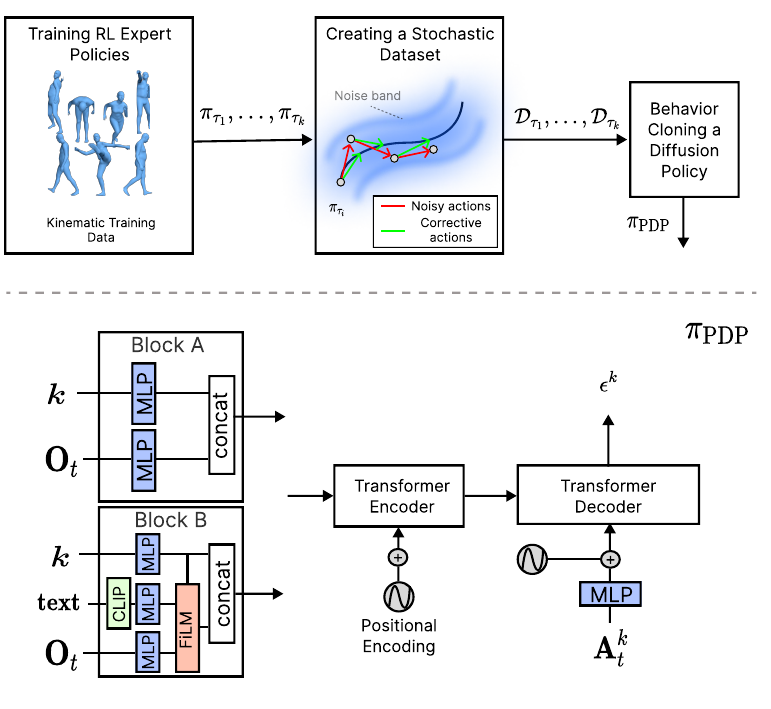}
\caption{ PDP Overview. Top: First, we train expert RL policies $\mathcal \pi_{\mathcal T_i}$ on tasks $\mathcal T_i$. We use $\mathcal \pi_{\mathcal T_i}$ to create a dataset of noisy-state clean-actions. We then use BC to train a diffusion model. Bottom: Our model is a transformer encoder-decoder architecture. Block-B is used for text-conditioned applications, while other applications use Block A. Note that these applications are trained separately on their own distilled dataset.}
\label{fig:arch}
\end{figure}

\subsection{Behavior Cloning with Diffusion Policy}
We parameterize our policy as a diffusion model. Here we give an overview of the diffusion model and our choice of architecture.


\subsubsection{Diffusion model}
We employ Diffusion Policy \cite{chi2023diffusionpolicy}, which models the action distribution conditioned on the observations as a denoising process using a Denoising Diffusion Probabilistic Model (DDPM) \cite{ho2020ddpm}. Given a dataset of sequences $\mathcal{D}$ collected using the method described previously, the denoising process is learned by a noise-prediction network $\epsilon_{\theta}(\bm{A}_t^k, \bm{O}_t, \bm{\tau}_t, k)$, where $\bm{A}_{t}^k$ is an action sequence sampled from $\mathcal{D}$ with added Gaussian noise and $k$ is the diffusion step. The diffusion model is conditioned on $\bm{O}_{t}$, the corresponding observation sequence. $\tau$ is the necessary task or goal information, and $\theta$ is the set of learned model parameters. As in Diffusion Policy, $\bm{A}_t^k$ is a length $T$ sequence of actions beginning at timestep $t$ and $\bm{O}_t$ is a length $T$ sequences of observations ending at timestep $t$. Sampling is then achieved through a denoising process known as Stochastic Langevin Dynamics \cite{welling2011bayesian} starting from pure random noise. 
\begin{equation}
    \bm{A}_{t}^{k-1} = \alpha (\bm{A}_{t}^k - \gamma \epsilon_{\theta}(\bm{A}_{t}^k, \bm{O}_{t}, \bm{\tau}_t, k) + \mathcal{N}(0, \sigma^2 \bm{I})),
\end{equation}

\noindent
where $\alpha, \gamma$ and $\sigma$ are hyper-parameters of the denoising process. The noise-prediction model is learned in a self-supervised manner using the mean squared error objective
\begin{equation}
    \mathcal{L} = MSE(\epsilon^k, \epsilon_{\theta}(\bm{A}_{t}^0 + \epsilon^k, \bm{O}_{t}, \bm{\tau}_t, k)),
\end{equation}

\noindent
where $\epsilon^k$ is the noise applied at to the action sequence at diffusion step $k$ and $\bm{A}_{t}^0$ is the clean action sequence.

\subsubsection{Model Architecture}
\label{sec:architecture}

Figure \ref{fig:arch} depicts our model architecture. We adopt a similar architecture to the time-series diffusion transformer proposed by Diffusion Policy \cite{chi2023diffusionpolicy}, with slight modifications depending on the application. For locomotion control and motion tracking, task information is included in the observation. For text-to-motion, the conditioning for the action sequence is computed as follows: a raw text prompt is encoded using the CLIP ViT-B/32 model \cite{CLIP} , then passed through an MLP text encoder. The observation $\bm{O}{t}$ is also fed through an MLP encoder. The diffusion step $k$ is embedded into the same space and added to the text embedding. This result is fed through a Feature-wise Linear Modulation (FiLM) layer \cite{perez2017FiLM}, which applies a learned element-wise scale and shift transformation to the embedding of $\bm{O}{t}$. Finally, the diffusion step embedding is concatenated with the FiLM layer result to produce our condition, serving as the input to the transformer encoder. This conditioning method is represented in Block B of Figure \ref{fig:arch}. The transformer decoder then takes an embedding of the noisy action sequence $\bm{A}_{t}^k$ along with the encoder result and predicts the noise applied to the action $\epsilon^k$. For motion tracking tasks, we condition our transformer encoder using Block A of Figure \ref{fig:arch}, which is similar to Block B but without the CLIP text.


\section{Experiments}
\label{sec:experiments}

%% file: experiments.tex
We demonstrate the generality and effectiveness of PDP by applying it to three distinctive applications using three different datasets: locomotion control under large physical perturbations using addbiomechanics dataset \cite{addbio}, universal motion tracking using AMASS \cite{amass2019}, and physics-based text-to-motion synthesis using the KIT subset of AMASS and HumanML3D \cite{guo2022Humanml3d} for text labels. Experimental details for each application are described in this section, with results, comparisons, and ablation studies presented in the next section.   Note that although each application uses the same model architecture in Figure \ref{fig:arch} (except for Block A and B), they are trained separately with different distilled datasets.



\subsection{Perturbation Recovery}
The goal of the perturbation recovery task is to train a single diffusion policy that is capable of capturing the wide range of human responses to perturbations. Being able to model and simulate this behavior is important for studying human robustness \cite{pushsim} and for designing better exo-skeleton or prosthetic systems \cite{hodossy2023shared}.

\subsubsection{Dataset}
We use the Bump'em dataset, a subset of the Addbiomechanics dataset \cite{addbio}, which consists of recovery motions of human participants being physically pushed while walking on a treadmill. Participants are perturbed in the same stance with varying forces and directions applied to the hip through a parallel tethered robot \cite{bumpem}. The recorded motions demonstrate that participants exhibit a diverse set of recovery strategies even under the same perturbation and initial stance, which makes this dataset particularly well suited for studying how well a model can capture the multi-modality of human behaviour. 

For this task, trials from one participant were used. Perturbations were collected as the subject walked forward on a treadmill at a fixed speed, impacted at left toe-off stance in four directions (front, left, right, back) with magnitudes of $7.5\%$ and $15\%$ body weight, resulting in 16 total motions; two were dropped due to poor motion quality.

\subsubsection{Experimental Details}
We use a 25-joint skeletal model from Addbiomechanics \cite{addbio}, optimized for the specific participant's inertia and joint lengths. The environment is simulated with Mujoco and consists of the simulated treadmill and the skeletal model. In this environment, the observation space is defined in the world frame. The RL agent's observation consists of the bodies center of mass positions $\bm{x}^{p} \in \mathbb{R}^{3B}$ and linear velocities $\dot{\bm{x}}^{p} \in \mathbb{R}^{3B}$, as well as the bodies rotation ${\bm{x}}^r \in \mathbb{R}^{3 \times 3 \times B}$  During RL training, the agent receives the same perturbation experienced by the human during the trial and optimizes tracking the human response through the collected reference motion following \cite{2018-TOG-deepMimic}. 

After training expert RL policies for each motion, we collect new observations for PDP, by including a binary signal $\bm{p}$ that indicates if the human is being perturbed, without detailing the force magnitude or direction.  This signal is helpful for the diffusion policy to differentiate between normal walking and perturbation recovery; otherwise, the policy would react to non-existent perturbations. This addition is justified, as humans can discern disturbances directly. Furthermore, this inclusion provides no predictive advantage, similar to providing foot contact information. The full perturbation recovery observation can then be defined as a tuple $(\bm{x}^p, \bm{x}^r, \dot{\bm{x}}^p, \dot{\bm{x}}^r, \bm{p})$. 15 motions were sampled using the various noisy/clean state action strategies, and used to train PDP. Each model was trained on a single RTX 2080 Ti GPU for approximately 1.5 hours.

\renewcommand{\arraystretch}{1.0} 
\begin{table*}[t]
    \centering
        \caption{Performance with different sampling strategies. The Tracking task was trained on the KIT subset of the AMASS train set and evaluated on the AMASS test set. The Perturbation task was trained on the Bump'em dataset. *Indicates our method.}
    \begin{center}
    \scriptsize 
    \resizebox{\textwidth}{!}{
        \begin{tabular}{c c | c c c c c | c c}
          \toprule 
            \multicolumn{2}{c|}{\textbf{Sampling Strategy}} & \multicolumn{5}{c|}{\textbf{Tracking Task}} & \multicolumn{2}{c}{\textbf{Perturbation Task}} \\
            \midrule
            \textbf{State} & \textbf{Action} & Success (\%) \(\uparrow\) & \(E_{\text{g-mpjpe}}\) \(\downarrow\) & \(E_{\text{mpjpe}}\) \(\downarrow\) & \(E_{\text{vel}}\) \(\downarrow\) & \(E_{\text{acc}}\) \(\downarrow\) & Success ID (\%) \(\uparrow\) & FPC \(\downarrow\)  \\
            \hline
            Clean & Clean & 68.8 & 57.1 & 33.1 & 8.77 & 5.55 & 3.36 & - \\
            Noisy & Noisy & 64.5 & 61.7 & 41.4 & 16.3 & 17.6 & 59.5 & 7.94 \\
            \textbf{Noisy*} & \textbf{Clean*} & 93.5 & 49.9 & 31.6 & 8.25 & 5.55 & 100.0 & 2.77 \\
            \bottomrule
        \end{tabular}
    }
    \end{center}

    \label{tab:sample_strategy}\vspace*{-9pt}
\end{table*}

\subsection{Universal Motion Tracking}

The goal of the universal motion tracking task is to train a single diffusion policy capable of controlling the character to track any given reference motion under physics simulation.

\begin{table*}[t]
\centering
\caption{Motion tracking results on AMASS train and test datasets.}
\begin{center}
\resizebox{\linewidth}{!}{
\begin{tabular}{c|c c c c c|c c c c c}
  \toprule
    \multicolumn{1}{c}{}  & \multicolumn{5}{c}{\textbf{AMASS-Train*}} & \multicolumn{5}{c}{\textbf{AMASS-Test*}} \\
    \midrule
    Method & Success (\%) \(\uparrow\) & \(E_{\text{g-mpjpe}}\) \(\downarrow\) & \(E_{\text{mpjpe}}\) \(\downarrow\) & \(E_{\text{vel}}\) \(\downarrow\) & \(E_{\text{acc}}\) \(\downarrow\) &  Success (\%) \(\uparrow\) & \(E_{\text{g-mpjpe}}\) \(\downarrow\) & \(E_{\text{mpjpe}}\) \(\downarrow\) & \(E_{\text{vel}}\) \(\downarrow\) & \(E_{\text{acc}}\) \(\downarrow\) \\
    \hline 
    MLP & 98.8 & 37.3 & 26.5 & 4.6 & 3.0 & 97.8 & 47.3 & 30.9 & 8.0 & 5.9 \\
    PHC & 98.9 & 37.5 & 26.9 & 4.9 & 3.3 & 96.4 & 47.4 & 30.9 & 9.1 & 6.8 \\
    PDP (Ours) & 98.9 & 36.8 & 26.2 & 4.7 & 3.3 & 97.1 & 46.2 & 30.2 & 8.0 & 5.7 \\
  \bottomrule
\end{tabular}
}
\end{center}
\label{tab:amass-mt}\vspace*{-9pt}
\end{table*}



\subsubsection{Dataset}
We use the AMASS dataset, and use the same train/test splits as PHC \cite{luo2023_PHC}, where the sub-datasets: Transitions Mocap and SSM synced are used for testing and the rest are placed in the training set. Since our method is agnostic to the RL policies themselves, we utilize pre-trained motion tracking controller PHC \cite{luo2023_PHC} to track most of the motions and train individual policies for challenging motions where PHC fails \cite{2018-TOG-deepMimic}. With PHC and a few specialized small policies, our training set covers most motions in AMASS \cite{amass2019} and KIT \cite{plappert2016kit}. Following the same practice in PHC, we exclude motions that are infeasible in our physics simulators, such as leaning on tables. 

\subsubsection{Experimental Details}
We use the humanoid model from \cite{luo2023_PHC}, which follows the SMPL kinematic structure with $J = 23$ spherical joints. The reference motion includes the linear position $\bm{x}^{p}_{ref} \in \mathbb{R}^{3J}$ and 6d rotation $\bm{x}^r_{ref} \in \mathbb{R}^{6J}$ of each joint. The linear and angular velocities are found by finite difference $\dot{\bm{x}}^p_{ref} \in \mathbb{R}^{3J}$, $\dot{\bm{x}}^r_{ref} \in \mathbb{R}^{3J}$ respectively. The full motion tracking observation can then be defined as a tuple $(\Delta\bm{x}^p, \Delta\bm{x}^r, \Delta\dot{\bm{x}}^p, \Delta\dot{\bm{x}}^r, \bm{x}^{p}_{ref}, \bm{x}^r_{ref})$, where $\Delta\bm{x}^p = \bm{x}^{p}_{ref} - \bm{x}^p$ and the other $\Delta$ terms are similarly defined. All quantities are measured in the character frame, where the origin is placed at the root of the character, the x-axis aligns with the character's facing direction, and the z-axis points upward.

We train PDP with a history of 4 observations and 1 action prediction horizon. For ablation experiments using KIT, we train on 4 \textit{NVIDIA A100} GPUs for approximately 24 hours. For experiments using AMASS, we train on 8 \textit{NVIDIA V100} GPUs for about 70 hours.

\subsection{Text-to-Motion} 
In the physics-based text-to-motion application, the goal is to train a diffusion policy to generate motions conditioned on a natural language text prompt.


\subsubsection{Dataset}
For training data, we use KIT dataset and the text annotations from HumanML3D \cite{guo2022Humanml3d}. The task vector is generated by passing the text annotation through the CLIP embedding \cite{CLIP}. We use the same pre-trained tracking controller, PHC, to obtain the observations and actions for training. 

\subsubsection{Experimental Details}
The observation space used to train PDP with the text-to-motion task is different from the tracking task. We use the joint position $\bm{x}^{p} \in \mathbb{R}^{3J}$ and linear velocities $\dot{\bm{x}}^{p} \in \mathbb{R}^{3J}$, as well as the joint rotation ${\bm{x}}^r \in \mathbb{R}^{6J}$ and rotational velocities $\dot{\bm{x}}^r \in \mathbb{R}^{3J}$. All quantities are measured in the character frame as defined in the motion tracking task. We use a history of 4 observations and 12 action prediction horizon. We train on 4 \textit{NVIDIA A100} GPUs for approximately 32 hours.

\section{Results} 
The experiments are designed to answer the following questions.
\begin{enumerate}[leftmargin=12pt]
\item Does the proposed sampling strategy, noisy-state-clean-action, outperform alternatives sampling strategies?
\item For the application of perturbation recovery during locomotion, can PDP achieve both robust and diverse control policy?
\item For universal motion tracking, how does PDP compare to the state-of-the-art RL policy and how important is it to use a generative model for this task?
\item Can we train a physics-based text-to-motion policy using PDP?  
\end{enumerate}


\subsection{Sampling Strategy}
We examine the impact of sampling strategies on the tracking and perturbation tasks. For the tracking task, training was conducted using the KIT dataset, with evaluations performed on the test set employed in the AMASS-Test split. Table \ref{tab:sample_strategy} records the quantitative results of different sampling strategies.

The clean-state clean-action approach has the lowest performance, with a success rate of  $3.36\%$ for perturbation and $68.8\%$ for tracking. In comparison, the noisy-state noisy-action strategy significantly improved the perturbation task to $66.9\%$, but the tracking performance dropped to $64.5\%$. A possible explanation for the drop in performance in the tracking task but increased performance in the perturbation task is the scale of datasets. The clean-state clean-action strategy restricted the perturbation dataset to just 14 unique trajectories, limiting its diversity. In contrast, the tracking task included 3,626 examples, making it less dependent on additional data from noisy sampling. Thus, random sampling strategies were likely more beneficial for the perturbation task as they introduced necessary variability lacking in the original dataset.

Noisy-state clean-action strategy outperforms other strategies, achieving a perfect success rate of $100\%$ in the perturbation task and $93.5\%$ success rate in the tracking task, see Table \ref{tab:sample_strategy}. These results underscore the importance of selecting the right sampling strategy. By visiting out-of-distribution states and collecting the optimal action, this approach generates higher quality demonstrations, leading to better generalization and robustness.

\subsection{Perturbation Recovery}
We compare the performance of PDP to two other baselines, a C-VAE and an MLP. The selection of these approaches was based on their reliance on supervised learning principles. The C-VAE introduces an alternative generative model framework, whereas the MLP serves as a deterministic alternative. The C-VAE uses a similar setup as \cite{characterControlVAE} where state and next state are fed into an encoder to produce a latent code. A decoder takes a randomly sampled latent vector alongside the current state to produce the current action. Both the C-VAE and MLP follow the same architecture as PDP with minor algorithm-dependent adjustments such as excluding the diffusion timestep embedding.

\subsubsection{Robustness}
Robustness is measured by the successful completion of an episode, defined as the agent not falling within 6 seconds of the perturbation, a period that allows for several gait cycles to complete post-perturbation. Perturbations are categorized into In-Distribution (ID), which are the same as those in the training data, and Out-of-Distribution (OOD) perturbations, which determines the policy's capacity to effectively handle unforeseen perturbations by adjusting aspects such as timing, intensity, and direction of impact. We sample OOD perturbations as follows: We choose a random perturbation to cover all gait phases by randomly choosing an impact timing within [0, 2] seconds (equivalent to ~2.5 gait cycles overlap), a random force magnitude between  $7.5\%$ and $15\%$ of body weight which represents the extrema of the forces used in the Bump'em dataset, and a random force direction that is parallel to the ground.

Table \ref{bumpem_baselines} shows the ID and OOD performance of each baseline. All three models can handle ID perturbations, achieving a success rate of $100\%$. However, when faced with OOD perturbations, C-VAE and PDP methods exhibit notably higher performance, with C-VAE achieving a success rate of $91.3\%$ and PDP achieving a success rate of $96.3\%$. Handling OOD impacts is challenging because the model may not have seen them before, especially the impact timing, as all training examples occur at the left toe-off. The OOD distribution performance results could be attributed to the multi-modal nature of the dataset, where using an MLP would result in policies that return the average of the response recorded in the dataset, causing the policy to fail, while C-VAE and PDP can synthesize a more tailored response from the multimodal distribution.

Two important hyper-parameters in our method are the choice of noise level for creating the stochastic dataset and the action prediction horizon. Note that the action prediction horizon refers to the number of actions being predicted during training. For inference, all policies re-plan after every step of the simulation. 
Table \ref{bumpem_ablation} shows the performance across different choices of noise level and horizon. Importantly, we see that 0 noise (equivalent to clean-state clean-action) performs extremely poorly with an ID success rate of just 3.36\%. Adding noise increases robustness significantly before eventually slightly harming performance. For the action prediction horizon, we note that lower horizons yield better performance, with horizon 1 achieving the best ID and OOD success rates of 100.0\% and 96.3\%, respectively. Larger horizons see a dramatic drop in performance. We speculate that this is because actions closer to the current timestep are more important for stability, but our loss function does not weight this importance.

\subsubsection{Foot Placement Correctness}

Foot placement holds significant importance in understanding perturbation response and balance \cite{perry2017walking, rebula2013measurement}. A reliable model should ideally mirror the human response or closely approximate it while also accounting for multi-modality. Figure \ref{fig:bumpem_distributions}a illustrates both the actual foot placements of the participant and the distribution of foot placements obtained through a noisy sampling procedure. 
When examining foot placement correctness for impacts within the distribution, a lower variance in foot placement should be anticipated. This is gauged by assessing the variance in the L2 distance between each foot placement and the nearest ground truth data point. 
We design a metric based on this called  foot placement correctness (FPC) to answer how spread apart the foot positions are from the policy compared to the closest ground truth position: 
\begin{equation}    
    \text{FPC} = \frac{1}{N} \sum_{i=1}^{N} \left( \min_{j \in \{1, 2, \ldots, M\}} \sqrt{(x_i - \bar{x}_j)^2 + (y_i - \bar{y}_j)^2}  \right)
\end{equation}
where $(x_i, y_i)$ refers to the foot contact position from a single policy rollout and $(\bar{x}_j,\bar{y}_j)$ refers to the ground truth foot contact positions.  

PDP achieves a much lower FPC score, $2.79$ compared to the top performing C-VAE model $\beta = .001$, $4.97$ as shown in Table \ref{bumpem_baselines}, signifying its proficiency in generating action state sequences that closely align with human responses. This is also demonstrated in Figure \ref{fig:bumpem_distributions} where it is clear that PDP aligns more closely to the ground truth foot contact positions than the C-VAE. 

C-VAE also struggles to capture multi-modality effectively. This is exemplified in the right impact, where the C-VAE prioritizes modeling one mode while PDP is able to generate both modes. Figure \ref{fig:bumpem_diffusion} shows two responses by the diffusion policy for the rightwards impact. Figure \ref{fig:bumpem_distributions} also displays the initial foot contact positions in response to the perturbation from both models. Balancing the reconstruction loss and the KL loss in C-VAE results in a significant trade-off in capturing the multi-modality and the variance in foot contact. Allowing the model to exhibit more variance enables better mode capture. Conversely, reducing the significance of the latent code leads to the model collapsing to a specific response with reduced variance. The comparison between Figure \ref{fig:bumpem_distributions}b and \ref{fig:bumpem_distributions}c may seem counter-intuitive in that increasing the $\beta$ term decreases variability. Nevertheless, this issue of posterior collapse is linked to C-VAE and is further discussed in the Discussion section.

\begin{table}[t]
    \centering
        \caption{Baseline comparisons on the Bump-em Dataset. ID stands for in-distribution, OOD stands for out-of-distribution, and FPC stands for foot placement correctness.}
        \begin{center}
        
    \resizebox{\columnwidth}{!}{
        \begin{tabular}{c | c | c c c}
          \toprule 
            Method & Beta Value & Success ID (\%) \(\uparrow\) & Success OOD (\%) \(\uparrow\) & FPC \(\downarrow\) \\
            \midrule
            \multirow{5}{*}{C-VAE} & 0.0001 & 98.1 & 91.0 & 5.09  \\
                                 & 0.001 & 100.0 & 91.3 & 4.97 \\
                                 & 0.01 & 100.0 & 71.0 & 4.26 \\
                                 & 0.1 & 99.8 & 61.3 & 5.27 \\
                                 & 1.0 & 99.8 & 59.0 & 5.28 \\
            \hline
            MLP & - & 100.0 & 81.0 & - \\
            PDP & - & 100.0 & 96.3 & 2.79 \\
          \bottomrule
        \end{tabular}
    }
    \end{center}

    \label{bumpem_baselines}\vspace*{-9pt}
\end{table}

\begin{table}[t]
\centering
\caption{Noise Level and Horizon Ablation for Bump'em Perturbation Task. FPC is measured on the left foot contact positions.}

\begin{center}

\resizebox{\columnwidth}{!}{
\begin{tabular}{c c | c c c }
  \toprule
    Noise Level & Horizon & Success ID (\%) \(\uparrow\) & Success OOD (\%) \(\uparrow\) & FPC \(\downarrow\)  \\
    \midrule
    0.0 & 6 & 3.36 & 0.0 & - \\
    0.08 & 6 & 97.9 & 75.0 & 2.26 \\
    0.12 & 6 & 100.0 & 90.0 & 2.77 \\
    0.16 & 6 & 90.5 & 83.0 & 2.98 \\
    \hline
    0.12 & 1 & 100.0 & 96.3 & 2.79 \\
    0.12 & 6 & 100.0 & 90.0 & 2.77 \\
    0.12 & 9 & 65.0 & 19.7 & 3.40 \\
    0.12 & 12 & 6.5 & 3.6 & - \\
  \bottomrule
\end{tabular}
}
\end{center}
\label{bumpem_ablation}\vspace*{-9pt}
\end{table}

\subsection{Motion Tracking}

We demonstrate that PDP is capable of reliably tracking a significant portion of the motion in AMASS. Our method achieves a $96.4\%$ success rate on the AMASS* test dataset, where failure is defined as in PHC~\cite{luo2023_PHC}: an episode is considered a failure if at any point during evaluation the joints are, on average, more than 0.5 meters from the reference motion. In addition to the success rate, we also adopt metrics used by \cite{luo2023_PHC}, specifically mean per-joint position error  ($E_{\text{mpjpe}}$) and global mean per-joint position error ($E_{\text{g-mpjpe}}$), which assess the model's accuracy in matching the reference motion in local and global frames. We also measure the error in velocity ($E_{\text{vel}}$) and acceleration  ($E_{\text{acc}}$) between the simulated character and the motion capture data. As shown in Table \ref{tab:amass-mt}, PDP matches or outperforms PHC in all metrics. 

Given the same architecture, we also compare with a regression model using MLP without diffusion, and find that MLP outperforms both PDP and PHC. This result is not entirely surprising because the benefit of generative models for the motion tracking application is not obvious in this task, as the action for tracking a particular reference pose from a particular state is not multi-modal but could be used as a motion prior.

\subsection{Text-to-Motion}

Figure \ref{fig:teaser} (middle) shows our results in the text-to-motion domain. We demonstrate that PDP is capable of following diverse text commands in natural language, such as jumping and kicking commands. Evaluating a model that employs auto-regressive inference during simulation with a limited history for composite actions presents significant challenges. Specifically, a prompt such as "walk then jump" cannot be effectively executed because the agent lacks the necessary memory of the initial action. To address this, we evaluate the model using a set of 42 action text prompts from the dataset, such as "a person dances" and "a person walks forward." Each prompt is considered successful if the model performs the specified action without falling. 

Diffusion models excel in handling multi-modal distributions, which may not significantly benefit the motion tracking task. However, in the text-to-motion application, where capturing multi-modality is crucial, diffusion models (PDP) significantly outperforms MLP  achieving a success rate of $57.1\%$ compared to $11.9\%$, respectively.

%% file: discussion.tex
\section{Discussion}


\paragraph{Robust Locomotion Policies}
Given the recent robotics community's interest in developing robust humanoid locomotion policies, our findings are particularly relevant. \cite{kaymak2023development, li2021reinforcement,singh2023learning}. While a single optimal strategy might suffice for a specific impact or perturbation, our results indicate that an MLP, which cannot capture different modes, lacks robustness to out-of-distribution (OOD) perturbations. In contrast, our diffusion model effectively stores a variety of strategies, providing it with a broader base of information to draw from when an OOD impact occurs. This capability could enhance locomotion policies' ability to handle diverse and unpredictable real-world perturbations.

\paragraph{C-VAE Posterior Collapse in Perturbation Recovery Task}Tuning the Beta value in C-VAE models presents significant challenges, primarily due to the posterior collapse problem. Increasing the $\beta$ value forces the latent distribution to align more closely with a normal distribution and can cause the model to disregard the latent vector and rely solely on the observation, effectively reducing the model to function like an MLP. We find that diffusion models cover the distribution of initial foot contact positions more effectively while requiring less tuning, making this model a preferred choice.

\paragraph{PDP and MLP Tracking Task}
Previous literature has shown the difficulty of producing a single and reliable motion tracker \cite{scalable-jessica, luo2023_PHC}. Our method can train a model directly through supervised learning and exceed the performance of more complex hierarchical RL policies. Furthermore, This capability facilitates the creation of a pre-trained tracking controller that can be swiftly adapted to new datasets. This feature is notably distinct from conventional RL-based methods \cite{scalable-jessica, won2020scalable, luo2023_PHC} which necessitates finetuning low-level controllers or training completely new ones.  Additionally, as hierarchical systems accumulate new low-level controllers, the composer faces increasing complexity, often necessitating a complete system retrain to reduce the number of low-level controllers. In contrast, our method only requires standard finetuning procedures to integrate a new dataset after training the local experts, enabling more efficient scalability with additional motion datasets.


\paragraph{Text2Motion Challenges}
Our system's capability extends to the text-to-motion task, demonstrating smooth transitions between text prompts despite a limited range of transitions.We hypothesize this effectiveness is partly due to the noisy sampling approach. By creating a band around each clean trajectory, we inadvertently increase the likelihood of intersecting with the state of another motion, facilitating more transitions.

However, our empirical observations indicate that text-to-motion does not perform at the same level as kinematic motion generation models. This discrepancy can likely be attributed to several factors. First, the model must balance performing the motion and maintaining equilibrium.  When losing balance, it compensates with small corrective steps, disrupting the original motion. Secondly, while combining two distinct motions can be close in kinematic space, like superimposing root rotation onto a jump to create a jump-and-turn motion, achieving these motions may be significantly different in skill space. That is, the agent faces a challenge in determining how to manipulate the feet to rotate the root while simultaneously executing the jump. 

\paragraph{Limitations}
Despite its advantages, our approach has notable limitations. The primary constraint is the speed of the denoising process.   Compared to the inference time of the MLP baseline, the diffusion model takes $K$ times longer, where $K$ is the number of denoising steps. This can make it challenging for applications that require high frequency control. Recent methods can reduce the number of inference steps required \cite{yin2023onestep, huang2024diffuseloco}. 

Another limitation lies in the trade-off inherent in diffusion-based policies when predicting over longer horizons. Although capturing multi-modality necessitates considering future actions, focusing on predicting multiple actions and weighting them equally can dilute the importance of the immediate action, thereby reducing robustness. This tension between long-horizon prediction for diversity and the accuracy of immediate actions is a crucial challenge. Future work could explore adaptive weighting schemes during training that balance effective long-horizon prediction with the precision of immediate actions.

%% file: conclusion.tex
\section{Conclusion and Future Work}

We present a novel framework for physics-based character control that leverages the capability of diffusion models to capture diverse behaviors. Our proposed sampling strategy, noisy-state-clean-action, significantly outperforms alternative sampling strategies. For the perturbation recovery task, our method effectively captures the distribution of human responses and demonstrates robustness to both in-distribution and out-of-distribution perturbations. In universal motion tracking, our method surpasses the state-of-the-art performance, including our baseline using a non standard non-generative model. Additionally, we showcase our methods ability to synthesize motion conditioned on text. Future work could look into speeding up the inference by using methods like  \cite{yin2023onestep, gu2023mamba}, or leveraging the large pre-trained motion tracker for downstream RL tasks.

%% file: figure_only.tex
\begin{figure*}[t]
    \centering
    \begin{subfigure}[b]{0.95\textwidth}
        \centering
        \includegraphics[width=\textwidth]{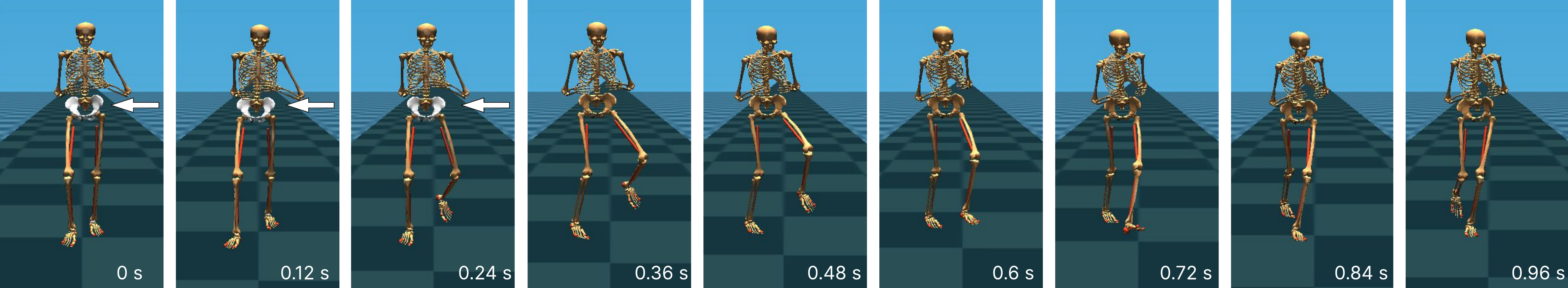} 
        \label{fig:subfig1}
    \end{subfigure}
    \vfill
    \begin{subfigure}[b]{0.95\textwidth}
        \centering
        \includegraphics[width=\textwidth]{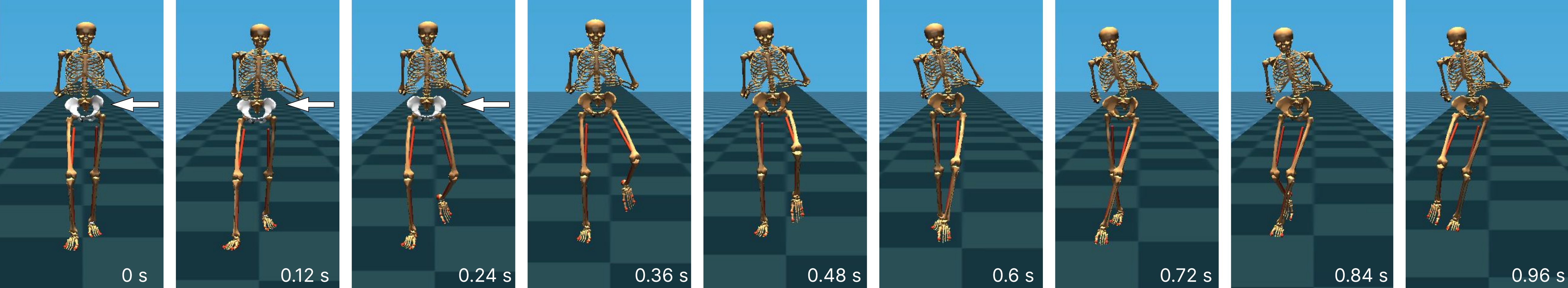} 
        \label{fig:subfig2}
    \end{subfigure}
    \caption{PDP rollouts for a 15\% body weight perturbation where the white pelvis and arrows indicate where the force is applied and the direction, respectively. Each row demonstrates a unique mode of recovery from the same perturbation.}
    \label{fig:bumpem_diffusion}
\end{figure*}

\begin{figure*}[h!]
    \centering
    \begin{subfigure}[b]{0.19\textwidth}
        \centering
        \includegraphics[width=\textwidth]{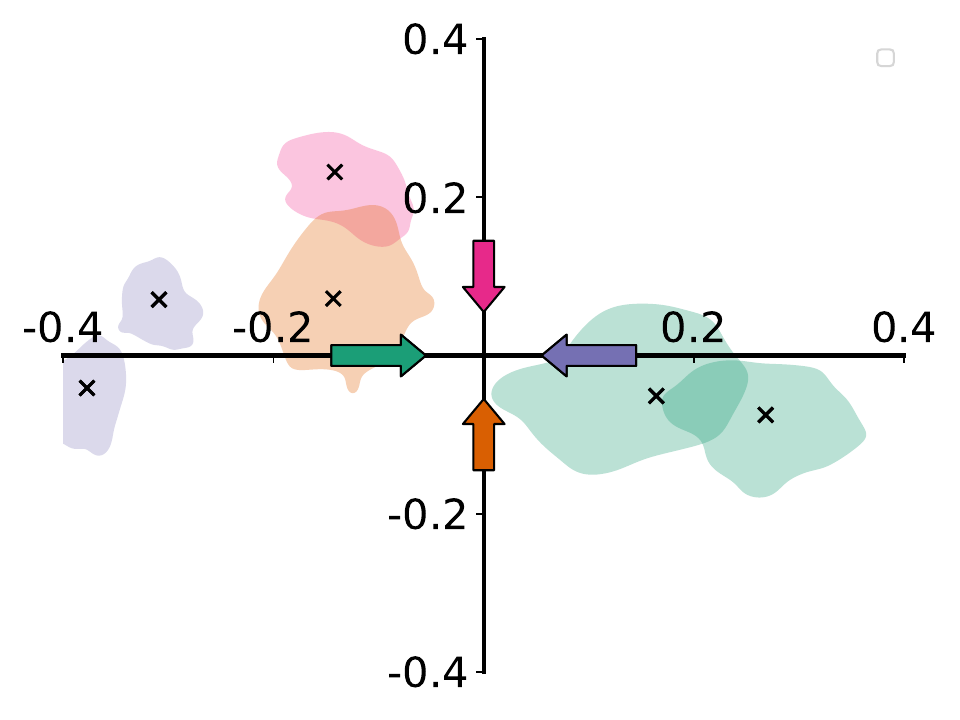} 
        \caption{Ground truth distribution}
        \label{fig:subfig1}
    \end{subfigure}
    \hfill
    \begin{subfigure}[b]{0.19\textwidth}
        \centering
        \includegraphics[width=\textwidth]{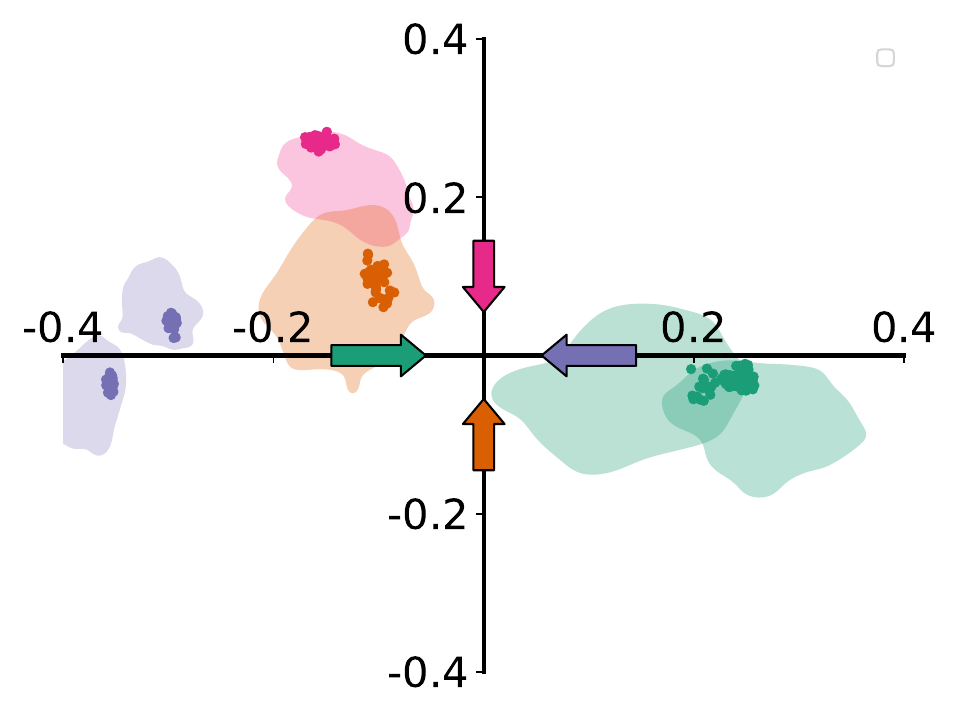} 
        \caption{VAE $\beta$ 0.01}
        \label{fig:subfig2}
    \end{subfigure}
    \hfill
    \begin{subfigure}[b]{0.19\textwidth}
        \centering
        \includegraphics[width=\textwidth]{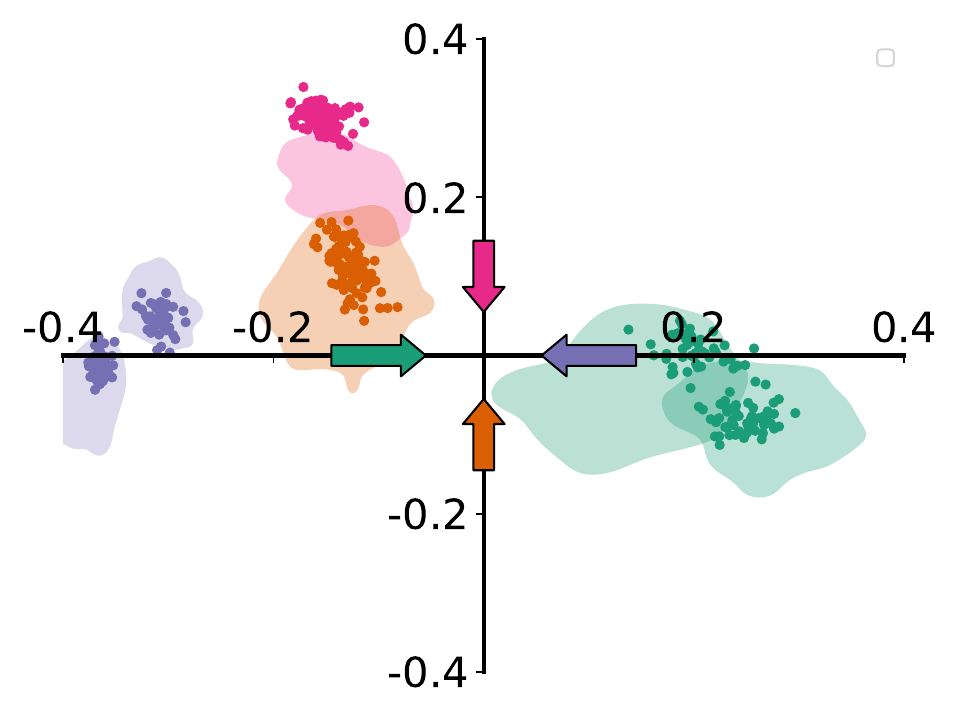} 
        \caption{VAE $\beta$ 0.001}
        \label{fig:subfig3}
    \end{subfigure}
    \hfill
    \begin{subfigure}[b]{0.19\textwidth}
        \centering
        \includegraphics[width=\textwidth]{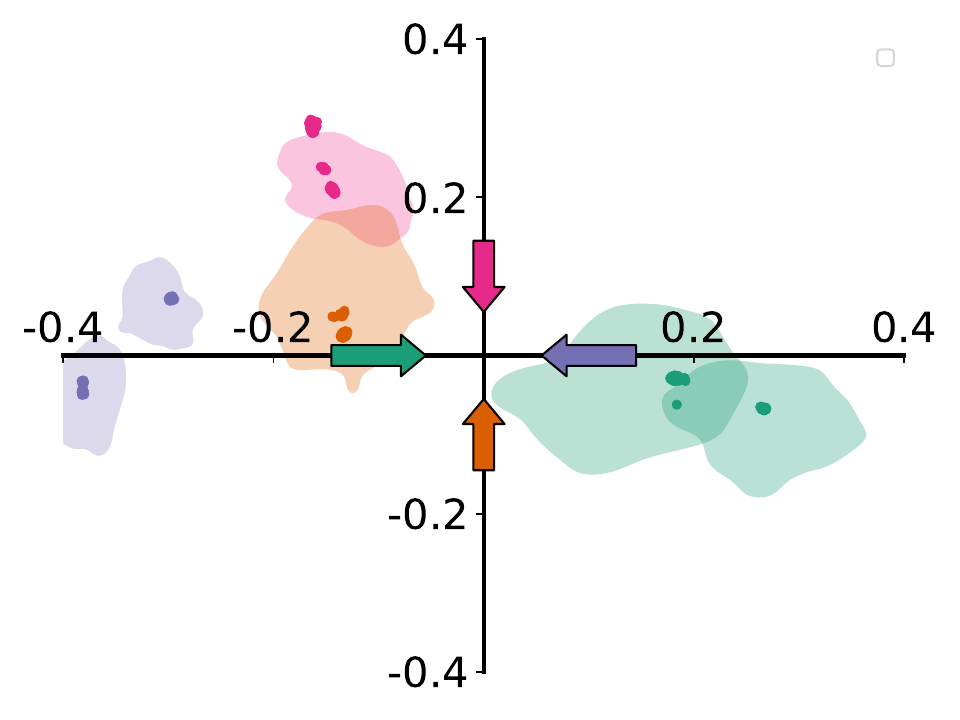} 
        \caption{PDP horizon 1}
        \label{fig:subfig4}
    \end{subfigure}
    \hfill
    \begin{subfigure}[b]{0.19\textwidth}
        \centering
        \includegraphics[width=\textwidth]{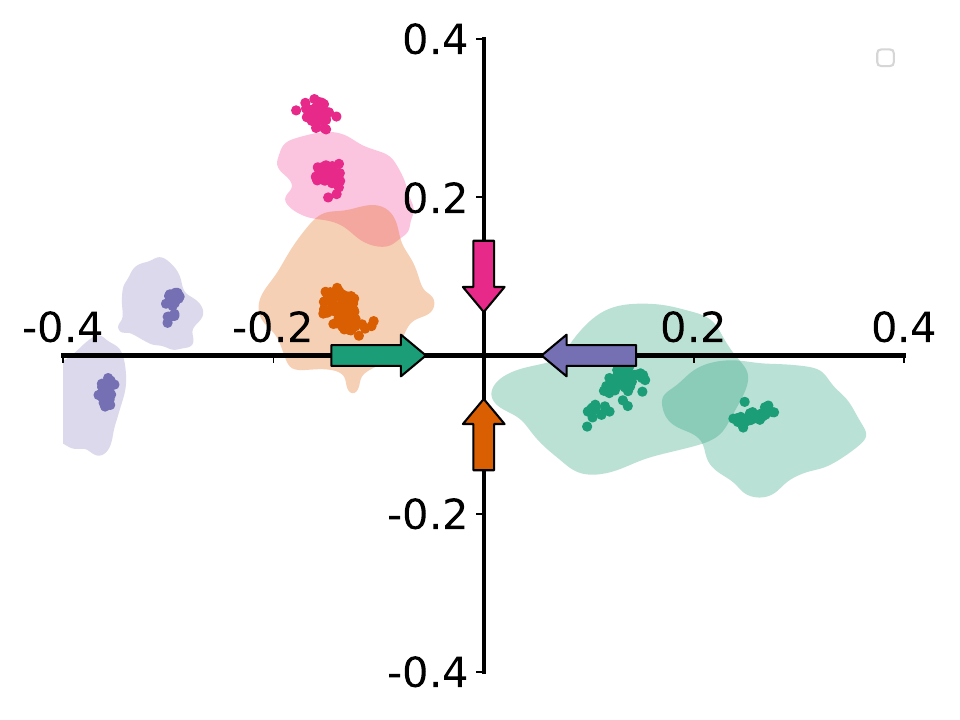} 
        \caption{PDP horizon 6}
        \label{fig:subfig5}
    \end{subfigure}
    \caption{Global left foot contact positions after $15\%$ body weight perturbation in meters. +Y and +X align with the character's forward and right directions, respectively. The different colored arrows represent the directions that the force is applied on the person. The shaded areas represent foot contacts in the training distribution with noise level 0.12. The black X's represent the ground truth foot contacts of the human participant. All policies were trained on the stochastic dataset with noise level 0.12.}
    \label{fig:bumpem_distributions}
\end{figure*}